\documentclass[aps,prb,10pt,
superscriptaddress,
reprint,
longbibliography]{revtex4-2}
\usepackage{hyperref}
\usepackage{graphicx}
\usepackage[utf8]{inputenc}
\usepackage{amsmath}
\usepackage{mathtools}
\usepackage{algorithm}
\usepackage{algpseudocode}
\usepackage{verbatim}
\usepackage{amssymb}
\usepackage{color}
\usepackage{mathtools}
\usepackage{bm}
\usepackage{stmaryrd}
\usepackage{subfigure}
\usepackage{tikz}
\usepackage{ifthen}
\usepackage{physics}
\usepackage{soul}
\usepackage{listings}
\usepackage[normalem]{ulem}
\hypersetup{
    colorlinks=true,
    linkcolor=blue,
    filecolor=gray,
    urlcolor=blue,
    citecolor=blue,
    }
\usetikzlibrary{calc}
\usepackage{cancel}

\definecolor{olivegreen}{HTML}{808000}

\def\etal{~\textit{et~al.}}

\begin{document}

\date{\today}\title{Uniaxial stress tuning of interfacial thermal conductance in cubic BAs/4H-SiC heterostructures}

\author{Lei Zhang}
\affiliation{Guangdong Provincial Key Laboratory of Magnetoelectric Physics and Devices, State Key Laboratory of Optoelectronic Materials and Technologies, Center for Neutron Science and Technology, School of Physics, Sun Yat-Sen University, Guangzhou, 510275, China}

\author{Fei Tian}
\affiliation{School of Materials Science and Engineering, State Key Laboratory of Optoelectronic Materials and Technologies, Sun Yat-sen University; Guangzhou, 510006, China.}

\author{Ke Chen}
\affiliation{Guangdong Provincial Key Laboratory of Magnetoelectric Physics and Devices, State Key Laboratory of Optoelectronic Materials and Technologies, Center for Neutron Science and Technology, School of Physics, Sun Yat-Sen University, Guangzhou, 510275, China}

\author{Zhongbo Yan}
\affiliation{Guangdong Provincial Key Laboratory of Magnetoelectric Physics and Devices, State Key Laboratory of Optoelectronic Materials and Technologies, Center for Neutron Science and Technology, School of Physics, Sun Yat-Sen University, Guangzhou, 510275, China}

\author{Kun Cao}
\email{caok7@mail.sysu.edu.cn}
\affiliation{Guangdong Provincial Key Laboratory of Magnetoelectric Physics and Devices, State Key Laboratory of Optoelectronic Materials and Technologies, Center for Neutron Science and Technology, School of Physics, Sun Yat-Sen University, Guangzhou, 510275, China}

\begin{abstract}
Understanding interfacial thermal transport is essential for improving thermal management in high-speed power electronic devices, where the efficient removal of excess heat is a critical challenge. In this study, a machine learning interatomic potential with near first-principles accuracy was employed to investigate the interfacial thermal conductance (ITC) between [111]-oriented cubic boron arsenide (cBAs) and [0001]-oriented 4H silicon carbide (4H-SiC), as well as its dependence on uniaxial stress. Among all possible bonding configurations at the cBAs(111)/4H-SiC(0001) interface, the B–C bonded interface was identified as the most energetically favorable. Non-equilibrium molecular dynamics simulations revealed that, under ambient conditions (300 K and 0 GPa), the ITC of the B–C interface reaches 353 $\pm$ 6 MW m$^{-2}$ K$^{-1}$, and increases monotonically to 460 $\pm$ 3 MW m$^{-2}$ K$^{-1}$ under a uniaxial stress of 25 GPa perpendicular to the interface. For comparison, the As–C bonded interface exhibits a lower ITC, increasing from 233 $\pm$ 7 to 318 $\pm$ 6 MW m$^{-2}$ K$^{-1}$ over the same stress range. These results demonstrate that proper interfacial bonding and moderate uniaxial stress can significantly enhance thermal transport across the cBAs(111)/4H-SiC(0001) heterointerface, offering valuable insight for thermal design in next-generation power electronics.
\end{abstract}

\maketitle
\section{Introduction}
Modern electronic devices demand superior thermal management to minimize Joule self-heating and ensure device stability and reliability  \cite{REview4TBCofWBG,zhang2025WBG}. To enable various electrical functionalities, these devices integrate multiple materials with distinct physical properties, inevitably giving rise to numerous interfaces. As heat generated during operation must traverse these heterogeneous boundaries, interfacial thermal conductance (ITC) emerges as a major bottleneck for enhancing overall heat dissipation efficiency. Therefore, understanding and tuning ITC is critical for improving thermal dissipation in modern electronic systems  \cite{RevModPhys94025002}. 

A natural approach is to employ materials with ultrahigh thermal conductivity for efficient dissipation of heat from localized hot spots. Compared to silicon, wide-bandgap semiconductors \cite{shur2019WBG,Ultrawide-BandgapSemiconductorsreview,WBG4cooling} offer higher breakdown voltages and saturation drift velocities, enabling greater integration density and reduced energy loss during high-frequency, high-voltage operation \cite{zhang2025WBG}. Among various wide-bandgap candidates, such as GaN and $\beta$-Ga$_2$O$_3$, 4H-SiC stands out with a thermal conductivity as high as around 415 W m$^{-1}$ K$^{-1}$ along the c-axis ([0001] orientation) \cite{4HSiCcalculatedTC,4HSiC-anisoTC,thermalConductanceofSiCPRM}, making it particularly well-suited for high-power, high-frequency and high-temperature electronic applications \cite{4HSiC-Cface,biela2010sicVsSi-based}. When integrated with high-thermal-conductivity (high-$\kappa$) substrates, 4H-SiC-based devices hold strong promise for enhanced thermal management, supporting their deployment in compact, high-power systems. 

Recently, the group III–V zinc-blende semiconductor cubic boron arsenide (cBAs) has been experimentally confirmed to exhibit the highest known isotropic thermal conductivity ($\sim$1300 W m$^{-1}$ K$^{-1}$) among all materials except diamond \cite{BAs-EXP1_li2018high,BAs-EXP2_kang2018experimental,BAs-EXP3_tian2018unusual}. Given the limitations of diamond materials, including high cost, limited wafer size, and challenges in achieving effective thermal bonding with other high-$\kappa$ materials \cite{Ultrawide-BandgapSemiconductorsreview,Diamond-shortcut_cheng2022high}, cBAs has emerged as a highly promising candidate for next-generation thermal management applications. Experimentally, an ITC of approximately 250 MW m$^{-2}$ K$^{-1}$ has been measured for a cBAs/GaN heterostructure bonded via an amorphous Al$_2$O$_3$ interlayer \cite{BAs/GaN_exp}. On the theoretical front, molecular dynamics (MD) simulations, which naturally account for high-order lattice anharmonicity, offer a significant advantage over conventional models such as the acoustic mismatch model \cite{AMM_model}, diffuse mismatch model \cite{DMM_model1,DMM_model2}, and atomistic Green's function \cite{AGF_1,AGF_2,AGF_3}, all of which typically neglect inelastic phonon scattering. The recent development of machine learning interatomic potentials (MLIPs) \cite{BPNN} has further expanded the applicability of MD in studying thermal transport \cite{MLonITC}. Using the Tersoff potential, Wei\etal \cite{BAs/Si} calculated the ITC of cBAs/Si interfaces with different crystallographic orientations, reporting values in the range of 200–300 MW m$^{-2}$ K$^{-1}$. More recently, Wu\etal \cite{BAs/GaN_Dp} developed a deep potential model \cite{DP} for the cBAs/GaN heterojunction based on MLIP, and their non-equilibrium molecular dynamics (NEMD) simulations predicted an ITC of $\sim$ 260 MW m$^{-2}$ K$^{-1}$ at 300 K for a direct interface without an interlayer. These findings underscore the considerable potential of cBAs in advanced thermal management applications.

Furthermore, applying mechanical stress to heterogeneous interfaces can continuously modulate lattice dynamics by altering the phonon properties of constituent materials and interfacial bonding strength \cite{kappaunderPressure,reversiblycontrollkappa}. In 4H-SiC-based power electronics \cite{sugiura2020piezoresistive4SiCdevices,sugiura2022high,StressOn4H-SiCdevices}, compressive stress can enhance carrier mobility via the piezoresistive effect \cite{piezoresistanceEffect}, thereby improving device performance. Notably, a lattice mismatch exists between the (111) plane of BAs (3.38 {\AA} \cite{BAs-EXP2_kang2018experimental}) and the (0001) plane of 4H-SiC (3.08 {\AA} \cite{elasticConstantsof4HSiC}), which exhibits the highest thermal conductivity among its orientations. This mismatch suggests that stress inevitably influences the ITC of cBAs/4H-SiC heterojunctions. However, research on the effects of mechanical stress on the ITC of cBAs/4H-SiC interfaces remains limited, with no experimental or theoretical studies addressing this topic to date.

In this work, a MLIP model based on the neuroevolution potential (NEP) framework \cite{NEP1,NEP2,GPUMDsoftware} is developed to enable the quantitative investigation of the ITC and its uniaxial stress dependence at the cBAs(111)/4H-SiC(0001) heterointerface. After validating the accuracy and transferability of this NEP model, the energetically favorable interfacial bonding configurations and optimal interlayer distances are identified, with the staggered B–C interface emerging as the most stable contact mode. Using NEMD simulations, the stress-dependent ITC of this interface is systematically investigated. At ambient conditions (300 K and 0 GPa), the B–C interface exhibits a high ITC of 353 $\pm$ 6 MW m$^{-2}$ K$^{-1}$. As the uniaxial stress increases to 25 GPa, the ITC rises approximately linearly to 460 $\pm$ 3 MW m$^{-2}$ K$^{-1}$. For comparison, the As–C interface, which is a metastable and less energetically favorable configuration than B–C interface, also shows a stress-induced enhancement in ITC, reaching 318 $\pm$ 6 MW m$^{-2}$ K$^{-1}$ at 25 GPa from 233 $\pm$ 7 MW m$^{-2}$ K$^{-1}$ at 0 GPa. This contrast further highlights the influence of interfacial bonding type on thermal transport.

\begin{figure*}[htbp]
    \centering
    \includegraphics[width=1\linewidth,keepaspectratio]{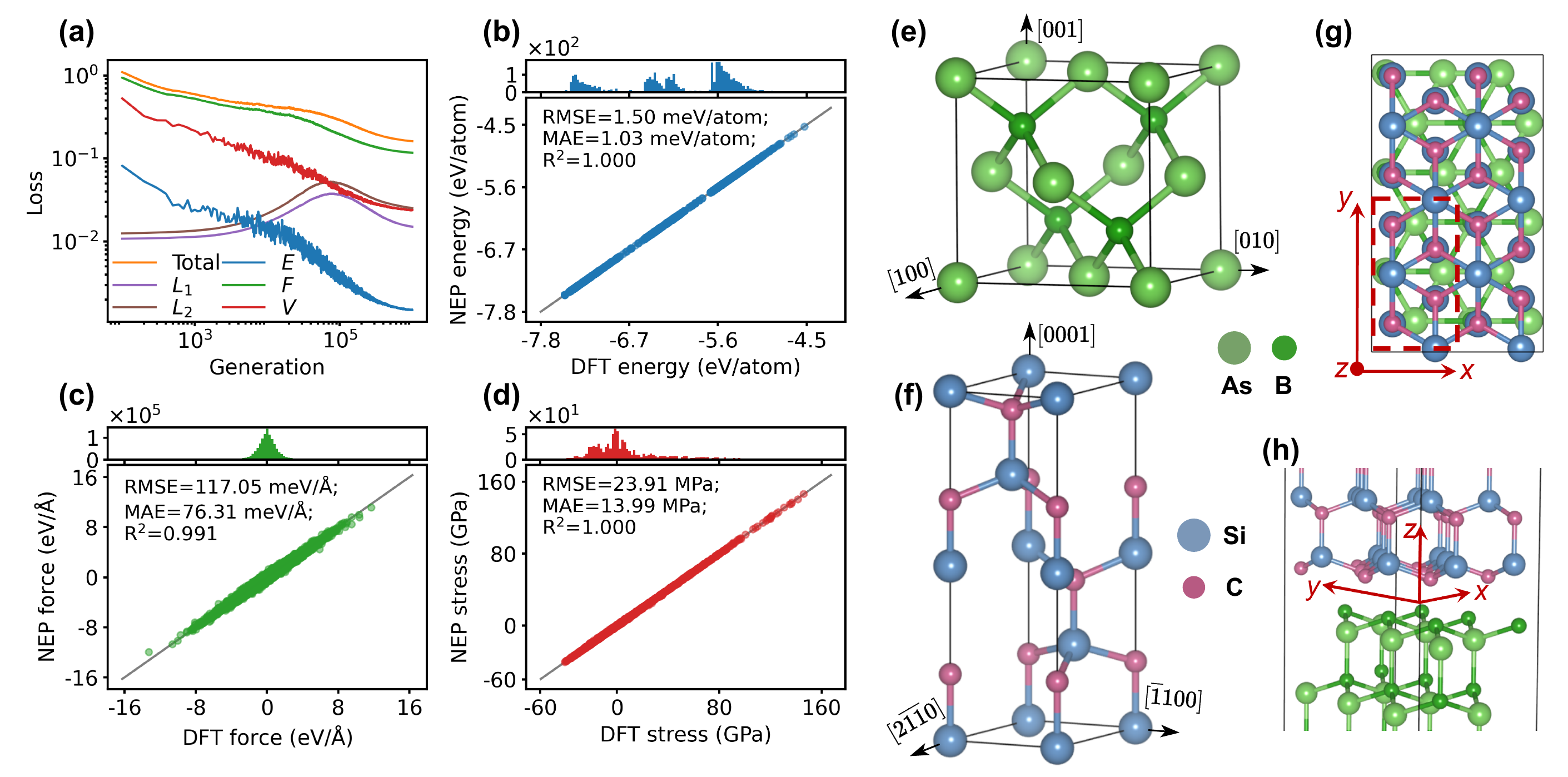}
	\caption{Regression evaluation of the NEP model and structures included in the training dataset. (a) Evolution of the total loss and individual loss components during NEP training. (b–d) Parity plots of NEP predictions versus DFT references for atomic energies, atomic forces, and virial stresses, respectively. Regression evaluation metrics including RMSE, mean absolute error (MAE), and R-squared (R$^2$) are presented in each panel. The corresponding data distributions are shown in the histograms above each subplot. Solid diagonal lines serve as visual guides. (e–f) Crystal structures of cBAs and 4H-SiC included in the dataset. (g–h) Interfacial structures featuring B–C bonding, shown from two different viewing angles, visualized using VESTA \cite{VESTA}. In (g), the side lengths of the red dashed rectangle equal half of the in-plane lattice parameters of the corresponding heterostructure.}\label{nepandstructure}
\end{figure*}
\section{Methods}

\subsection{NEP training and DFT calculations}

The NEP approach utilizes a separable natural evolution strategy for training, with an atom-environment descriptor-based artificial neural network architecture. The NEP model is trained to establish a mapping between the descriptor vector $q_{\nu}^i$ (with $N_\mathrm{des}$ dimensions) of the local environment of a given atom $\textit{i}$ and its environment-dependent site energy $U_i$:

\begin{equation}
	U_i\left(\{q_{\nu}^i\}_{\nu =1}^{N_\mathrm{des}}\right) = \sum_{\mu =1}^{N_\mathrm{neu}}w_{\mu}^{(1)}\tanh \left(\sum_{\nu =1}^{N_\mathrm{des}}w_{\mu\nu}^{(0)}q_{\nu}^i-b_{\mu}^{(0)}\right)-b^{(1)}, 
\end{equation}

where $\tanh(*)$ denotes the activation function at the hidden layer, and $N_\mathrm{neu}$ represents the number of neurons. The parameters $w_{\mu}^{(1)}$, $w_{\mu\nu}^{(0)}$, $b^{(0)}_{\mu}$, and $b^{(1)}$ are the trainable weights and biases in the neural network of the NEP.

The training process of the machine learning interatomic potential involves minimizing the loss function, which is typically related to the fitting targets in the training dataset. In the NEP approach, the loss function, $L= \lambda_e \Delta E + \lambda_f \Delta F + \lambda_v \Delta V + \lambda_1 L_1 + \lambda_2 L_2$, is designed as a weighted sum of the root-mean-square errors (RMSEs) between the predicted and density functional theory (DFT) reference values for energy ($\Delta E$), force ($\Delta F$), and virial ($\Delta V$), along with two regularization terms $L_1$ and $L_2$ to prevent overfitting. The coefficients $\lambda_e$, $\lambda_f$, $\lambda_v$, $\lambda_1$, and $\lambda_2$ in front of each term serve as weight factors in the loss function.

The dataset for this study was constructed following a similar procedure to that used in our previous work \cite{aluminaforworkflow}. Random strains (-10\% to 10\% for lattice scaling and -5\% to 5\% for lattice distortions) and moderate atomic displacements (0.1 {\AA}) were applied to supercells of various sizes for cBAs and 4H-SiC to sample their elastic responses, deformation behaviors, and thermodynamic vibrations. To reduce interfacial interactions and improve sampling efficiency, sufficiently large superlattice models containing two distinct interfaces were used in place of conventional heterostructure models. For interface structures, various interfacial contact types were constructed and random atomic displacements were applied to sample the atomic environments near the interfaces. Based on DFT calculations of these perturbed structures, an initial NEP model was trained and subsequently refined through active learning by iteratively sampling physically relevant configurations and expanding the training dataset until the final NEP model met the desired level of accuracy.

All structures contained in the dataset, along with their corresponding energies, atomic forces, and virial stresses, were computed using the Vienna Ab Initio Simulation Package (VASP) \cite{VASPsoftware}, employing the Projector Augmented-Wave (PAW) method \cite{PAW} and the Perdew–Burke–Ernzerhof (PBE) functional within the Generalized Gradient Approximation (GGA) \cite{GGAmadesimple}. The DFT calculations were performed with a cutoff energy of 500 eV, a $\Gamma$-centered k-spacing of 0.15 {\AA}$^{-1}$ to generate evenly spaced k-meshes, and a threshold of 10$^{-6}$ eV for the electronic self-consistent loop. More detailed information on dataset construction and the NEP model training can be found in the supplementary materials (SM) and Fig. S1 \cite{SM}.

\subsection{Construction of cBAs(111)/4H-SiC(0001) Interface}

The interface structures used in both the dataset and subsequent MD simulations were constructed as follows. First, the $\sqrt{3} \times \sqrt{3}$ (111) plane of cBAs  and the $2 \times 2$ (0001) plane of 4H-SiC were stacked in an upright orientation, resulting in a small lattice mismatch of $\sim$ 4.5\% between the cBAs (111) and 4H-SiC (0001) planes ($\sqrt{3} \cdot 3.40$ (cBAs) vs. $2 \cdot 3.09$ (4H-SiC) after structure relaxation in this study). The resulting supercell was then transformed into an orthogonal cell to facilitate ensuing calculations, yielding in-plane lattice constants of $\sqrt{3} \cdot 3.40$ and $3 \cdot 3.40$ {\AA} along the x and y directions, respectively, as illustrated in Fig. \ref{nepandstructure}(g-h).

\subsection{NEMD simulations}

\begin{figure}[htbp]
    \centering \includegraphics[width=1\linewidth,keepaspectratio]{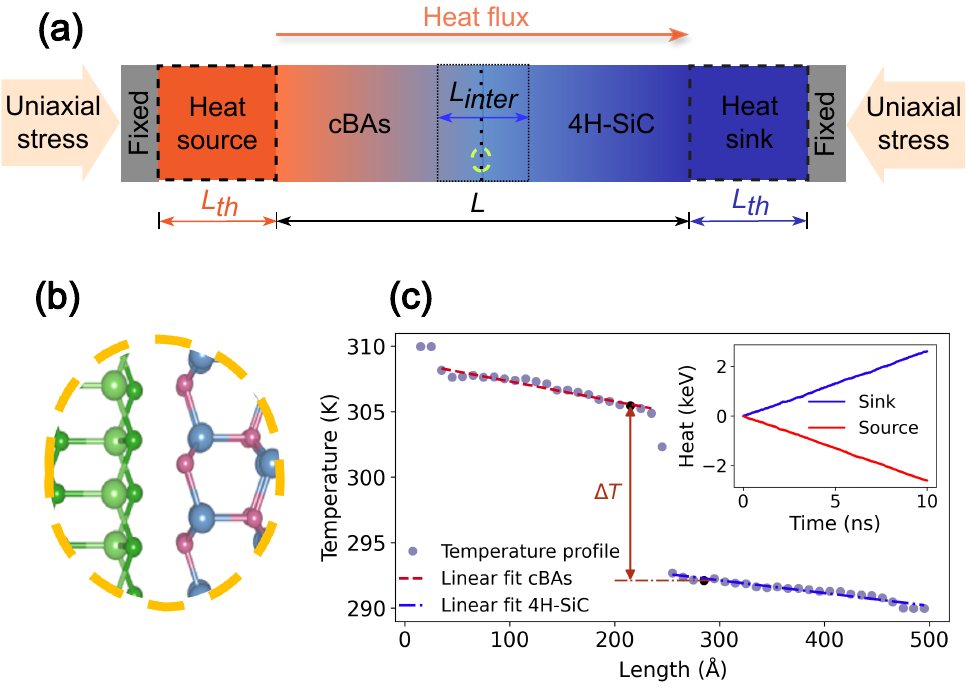}
    \caption{(a) Schematic diagram of the interfacial model setup used for NEMD simulations; (b) Magnified view of the dashed-circle region at the interface in (a), illustrating the atomic configuration with B–C interfacial bonding; (c) Temperature profile across the interface at 300 K and 0 GPa obtained from the NEMD simulation. The model was divided into 50 slices along the transport direction, each with a thickness of approximately 10 {\AA}, and the temperature of each slice was calculated to obtain the temperature profile. The black temperature points on both sides of the interface represent the reference points used to determine the actual temperature drop, $\Delta T$, in this study. The inset shows the cumulative energy exchanged by the thermostats coupled to the heat source and heat sink.}
    \label{tempprofile}
\end{figure}

The molecular dynamics simulations of ITC were performed using the NEMD method implemented in the GPUMD software (\lstinline{version 3.9.5}) \cite{GPUMDsoftware}. In NEMD, non-equilibrium steady-state heat flux is established in the structure by setting local thermostats at different temperatures, allowing the calculation of the ITC $G$:
\begin{equation}
    G=\frac{\langle J\rangle }{A\Delta T},
\end{equation}
where $\langle J\rangle$ is the average energy flux along the temperature gradient (the transport direction), $A$ is the cross-sectional area perpendicular to the transport direction, and $\Delta T$ represents the temperature drop across the interface as shown in Fig. \ref{tempprofile}(c). The ITC $G$ can be further decomposed to obtain the contribution of phonon modes with different frequencies to the spectrally decomposed ITC \cite{HNEMD}, $G(\omega)$:

\begin{equation}
    G=\int_{0}^{\infty}\frac{d \omega}{2\pi} G(\omega) ,
\end{equation}
where
\begin{equation}
    G(\omega) = \frac{2 }{A\Delta T}\int_{-\infty}^{+\infty}e^{i\omega t}K(t)dt. 
\end{equation}
Here, $K(t)$ is the virial-velocity-time correlation function \cite{SHC} along the transport direction, which in its full vector form is given by:

\begin{equation}
    \mathbf{K}(t) = \sum_i \langle \mathbf{W}_i(0)\cdot \mathbf{v}_i(t) \rangle,
\end{equation}
where $\mathbf{W}_i$ and $\mathbf{v}_i$ are the per-atom virial tensor and velocity vector of atom $i$.

The cBAs(111)/4H-SiC(0001) structure used in the NEMD simulations consists of a rectangular simulation cell containing 40,200 atoms, with in-plane dimensions of 29.5 {\AA} $\times$ 30.6 {\AA} and lengths of 501.8 {\AA} and 502.2 {\AA} along the transport direction for B–C and As–C bonding interface, respectively. To account for the size effects inherent in NEMD simulations, the system size was carefully tested to ensure the convergence of the computed ITC (Fig. S2 \cite{SM}).

As shown in Fig. \ref{tempprofile}(a), heat flux is imposed from the cBAs side toward the 4H-SiC side. Periodic boundary conditions were applied in all directions. To avoid unphysical heat recirculation due to periodicity, the outermost 15 {\AA} at both ends of the transport direction were fixed. Adjacent to these fixed regions are $L_{th}$=60 {\AA}-wide heat source and heat sink regions located within cBAs and 4H-SiC, respectively.

The simulation protocol began with a 1 ns relaxation using a Berendsen thermostat \cite{berendsenthermo} in the NPT ensemble, maintaining a temperature of 300 K and applying uniaxial stress along the transport direction when specified. This was followed by a 100 ps NVT relaxation using a Nos\'{e}-Hoover chain thermostat \cite{NHCthermo}. Subsequently, Langevin thermostats \cite{langevinthermo} were applied to the heat source and sink regions to maintain temperatures of 310 K and 290 K, respectively, for 1 ns. The system then evolved for an additional 10 ns to reach a steady state, during which temperature and energy profiles were collected over the final 5 ns.

A time step of 1 fs was used throughout the simulations. The damping parameter for temperature control was set to 0.1 ps, while for pressure control in the NPT stage, a damping constant of 1 ps was applied. Since uniaxial stress induces strain through the Poisson effect, the system dimensions may vary under different applied pressures. To maintain clarity and consistency, all reported dimensions refer to those of the initial (unstressed) configuration unless otherwise specified. For numerical evaluation of the ITC, however, the cross-sectional area corresponding to the deformed structure under each applied pressure was used.

To enable a fair comparison between the two types of interfaces, the temperature drop $\Delta T$ was defined based on the temperature profile measured in regions located 40 {\AA} away from the interface on both sides, corresponding to an interfacial region length of $L_{\mathrm{inter}}$ = 80 {\AA}, as illustrated in Fig. \ref{tempprofile}(a-c). For each simulation condition, five independent runs were performed to obtain averaged results and corresponding standard errors. 

\section{results}

\begin{figure*}[htbp]
    \centering
    \includegraphics[width=1\linewidth,keepaspectratio]{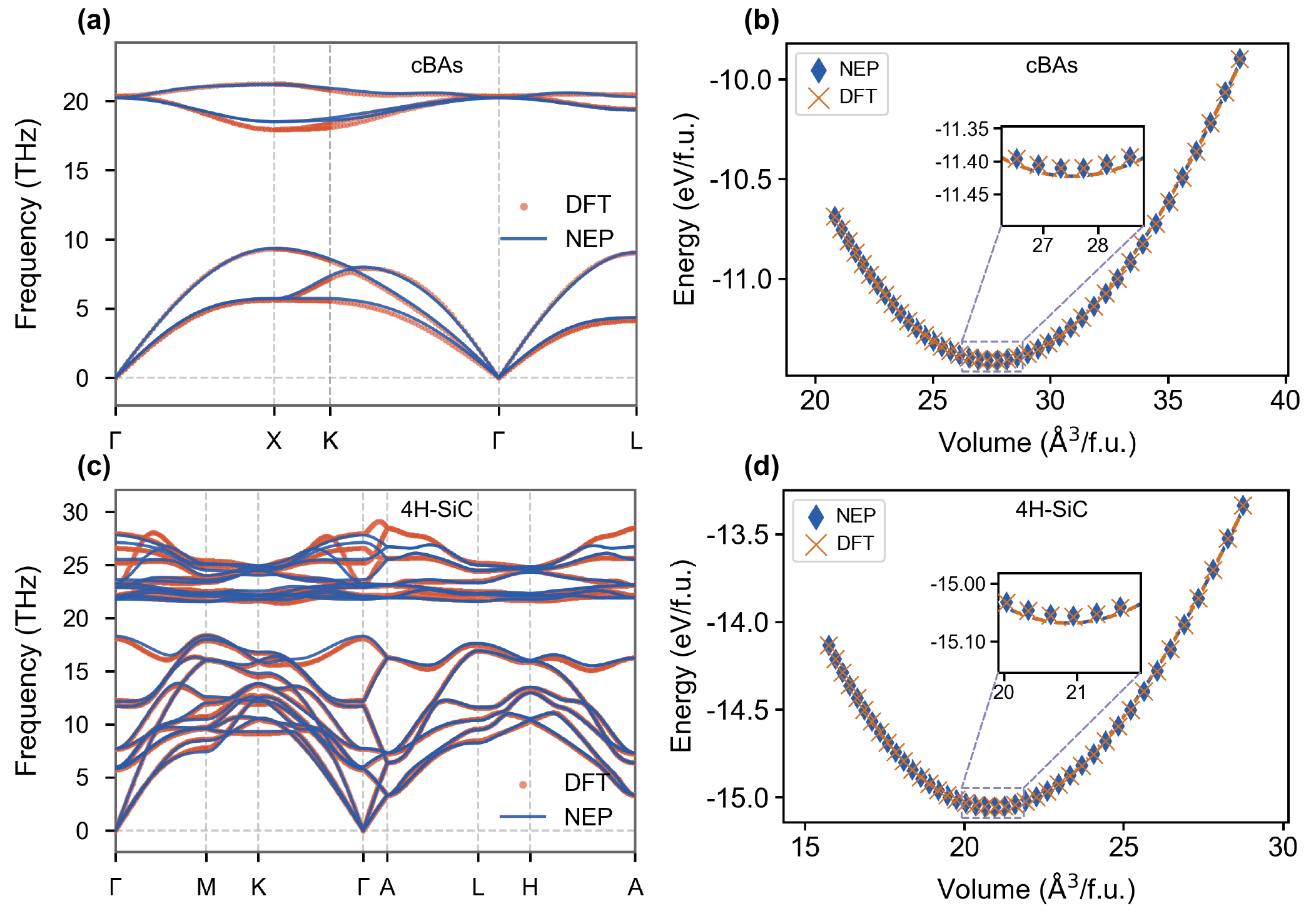}
	\caption{Phonon spectra and EOS of cBAs and 4H-SiC. Panels (a–b) and (c–d) show the results for cBAs and 4H-SiC, respectively. In the phonon plots, red circles represent DFT results, while blue solid lines indicate NEP predictions. In the EOS plots, blue diamonds and yellow crosses denote NEP and DFT data points, respectively. The blue solid and yellow dashed curves are the Birch–Murnaghan fits to the NEP model and DFT data, respectively. Insets highlight the details near the equilibrium volume.}
	\label{phonoAndEOV}
\end{figure*}

\subsection{Evaluation of the NEP model}
 
Based on structures selected through perturbation and an active learning procedure, a training dataset comprising 1,770 configurations was constructed. This dataset includes 810 pure cBAs and 420 4H-SiC structures of various sizes, along with 540 heterostructures involving four distinct interface types: As–C, B–C, As–Si, and B–Si. The NEP model employed in this work was trained on this dataset for one million steps. The left panel of Fig. \ref{nepandstructure} illustrates the training process, along with parity plots for atomic energies, forces, and virial stresses. The RMSEs are 1.50 meV/atom, 117.05 meV/{\AA}, and 23.91 MPa, respectively, indicating high training accuracy. The model also demonstrates strong generalization, as validated on a test set consisting of out-of-sample structures representative of our target application scenarios. The corresponding parity plots for the test set are provided in the SM \cite{SM}, showing similarly low prediction errors and suggesting that the relevant phase space has been adequately explored.

To further evaluate the NEP model, we compared its predictions for several static properties of the elemental materials cBAs and 4H-SiC with DFT results computed at a level of accuracy comparable to that used for dataset construction. These properties include phonon dispersion (see Fig. \ref{phonoAndEOV}(a) and \ref{phonoAndEOV}(c) for cBAs and 4H-SiC, respectively), equations of state (Fig. \ref{phonoAndEOV}(b) and \ref{phonoAndEOV}(d)), and elastic constants (see Table \ref{cij}). The NEP model shows good overall agreement with DFT for both materials. Minor discrepancies are observed in the optical branches of the phonon spectrum for 4H-SiC. This deviation is attributed to the longitudinal optical–transverse optical (LO–TO) splitting near the $\Gamma$ point, which arises from polarization induced by atomic displacements and the associated macroscopic electric field. Capturing this effect in first-principles calculations requires inclusion of non-analytical corrections. However, such long-range electrostatic interactions cannot be described by the current NEP framework, which lacks explicit treatment of charge and polarization degrees of freedom. Similar limitations have also been reported in other MLIPs applied to SiC polymorphs that do not incorporate these physical features \cite{xie2023uncertainty}.

The EOS predictions by the NEP model are in close agreement with DFT results. Consistent lattice constants are obtained for both materials: for zinc-blende cBAs, both NEP and DFT yield a = 4.81 {\AA}, and for 4H-SiC, a = b = 3.09 {\AA}, c = 10.11 {\AA}. Regarding elastic constants, the NEP model matches the DFT results well for cBAs, though both methods slightly underestimate the experimental stiffness. In contrast, the results for 4H-SiC show better agreement with experimental values. It is worth noting that experimental measurements inherently include phonon zero-point vibrations \cite{ZPAEs} and thermal effects, which are not captured by DFT or the NEP model. Additionally, the equilibrium lattice constant of cBAs from DFT (4.81 {\AA}) is slightly larger than the experimental value (4.78 {\AA} \cite{BAs-EXP2_kang2018experimental}), which may partly explain the observed discrepancy in stiffness.

\begingroup
%\squeezetable
\begin{table}[!htb]
	\begin{ruledtabular}
	\centering
	\caption{The elastic stiffness constants C$_{ij}$ (GPa) of cBAs and 4H-SiC, calculated by NEP and DFT, utilize stress-strain methods through VASPKIT \cite{VASPKIT}.}
	\label{elastic_constants}
	\begin{tabular}{llcccccc}
	materials & Method & C$_{11}$ & C$_{12}$ & C$_{13}$ & C$_{33}$ & C$_{44}$ & C$_{66}$ \\
	\hline
	cBAs & DFT & 266 & 64 & - & - & 145 & - \\
				& NEP & 268 & 64 & - & - & 145 & -\\
				& Exp \cite{elasticConstantsofc-BAs} & 291$\pm$5 & 75$\pm$13 & - & - & 173$\pm$6 & -\\
	4H-SiC & DFT & 512 & 103 & 55 & 552 & 162 & 205 \\
				& NEP & 503 & 111 & 55 & 548 & 160 & 196\\
				& Exp \cite{elasticConstantsof4HSiC} & 501 & 111 & 52 & 553 & 163 & -\\
	\end{tabular}\label{cij}
	\end{ruledtabular}
\end{table}
\endgroup

\begin{figure*}[htbp]
    \centering \includegraphics[width=1\linewidth,keepaspectratio]{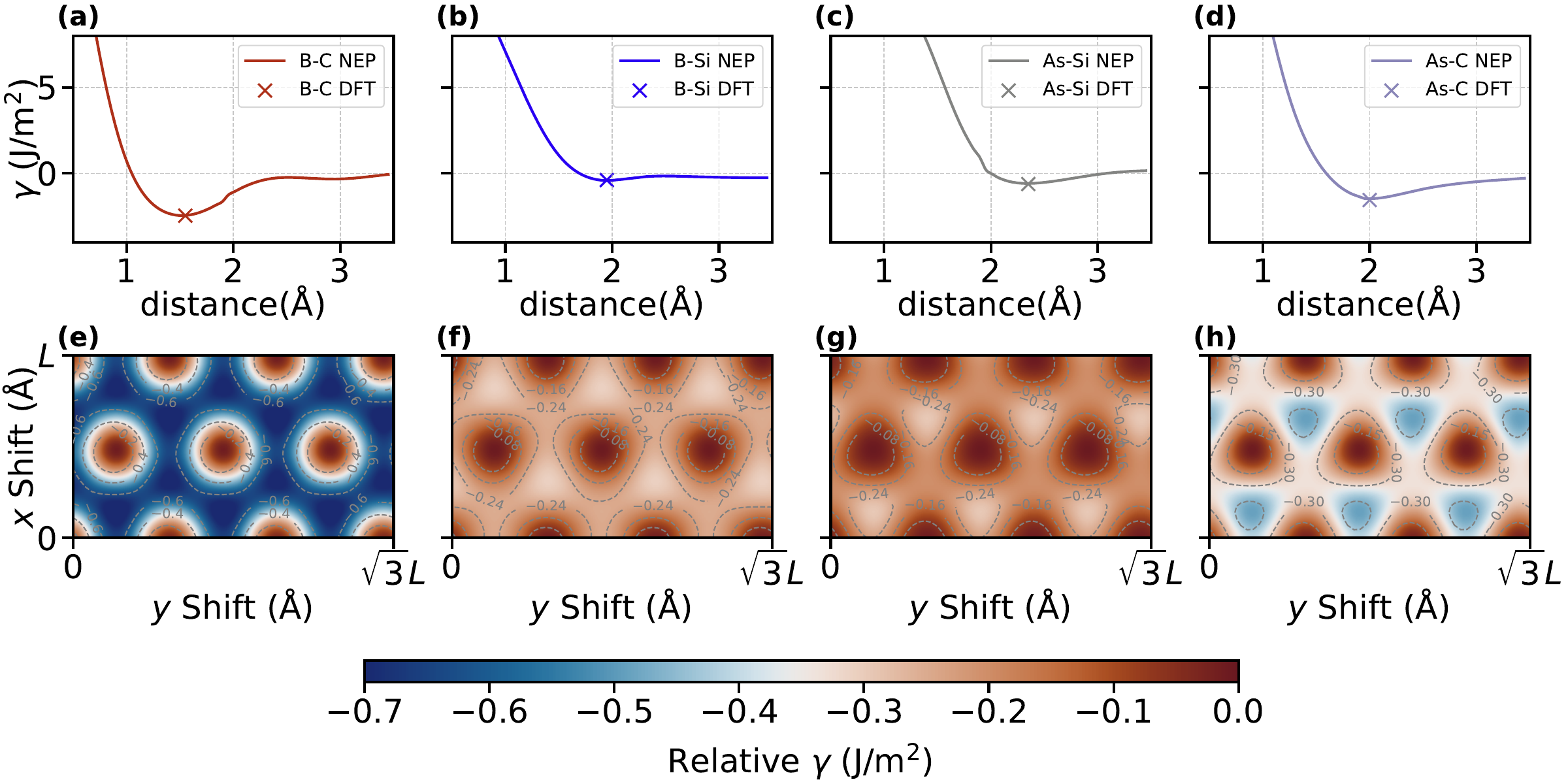}
	\caption{Interfacial energetics of cBAs(111)/4H-SiC(0001) with different atomic bonding configurations. (a–d) Interfacial energy as a function of interlayer spacing for interfaces formed by B–C, B–Si, As–Si, and As–C bonding, respectively. Cross markers indicate DFT-calculated energies at the energy-favored interfacial distances predicted by NEP model. (e–h) Interfacial energy landscapes corresponding to lateral sliding at the equilibrium interlayer spacing for each bonding type, illustrating the relative stability and shear resistance of the interface configurations. The maximum interfacial energy encountered along the sliding path is taken as the reference zero for each sliding interface. The region labeled as $L\times \sqrt{3} L$ in (e–h) corresponds to the red dashed rectangle area in Fig. \ref{nepandstructure}(g), i.e. $L = \frac{\sqrt{3}}{2}\cdot3.40 $ {\AA}.}
	\label{interfaceEnergyMap}
\end{figure*}
By incorporating atomic velocities from MD simulations, spectral energy density (SED) method \cite{comparisonSED,ThomasSED,EqualityofBothSED,larkin2014comparisonSED} enables the extraction of phonon properties such as dispersion relations and lifetimes under various thermodynamic conditions. To evaluate whether the NEP model accurately captures phonon evolution in cBAs and 4H-SiC under uniaxial stress, we applied the SED approach to analyze phonon behavior along the [111] direction of cBAs and the [0001] direction of 4H-SiC under applied stress. The results were compared against experimental data reported in Ref. \cite{li2022anomalous}, which include the pressure-dependent evolution of the acoustic–optical phonon gap, sound velocity, and maximum optical phonon frequency in cBAs. Our NEP model yields values and trends that are in good agreement with these experimental observations. The phonon evolution in 4H-SiC under stress also aligns well with expectations. See Fig. S3 and the SM for further details \cite{SM}. These results confirm that the NEP model reliably captures the stress-induced phonon responses of the constituent materials at finite temperature. 

\subsection{Interfacial bonding type }

For 4H-SiC crystals grown along the [0001] direction, the surface termination can consist exclusively of either silicon or carbon atoms, commonly referred to as the Si-face and C-face, respectively \cite{4HSiC-Cface}. Similarly, when cleaving cBAs along the [111] direction, the resulting (111) surface can be terminated by either boron or arsenic atoms, denoted as the B-face and As-face. When these differently terminated surfaces of 4H-SiC(0001) and cBAs(111) come into contact, four distinct interface bonding configurations are possible: B–C, B–Si, As–Si, and As–C.

Using the validated NEP model, the interface energy was calculated to evaluate the relative stability of these bonding configurations, defined as:
\begin{equation}
    \gamma = \frac{1}{A}\left[E_{tot}-(E_{\text{cBAs}}^{slab} + E_{\text{4H-SiC}}^{slab})\right],
\end{equation}
where $\gamma$ is the interface energy, $E_{tot}$ is the total energy of the simulation cell containing both slabs, and $A$ is the cross-sectional area of the interface. By varying the interfacial separation between the two slabs, the energy profile as a function of interface distance can be obtained, from which the equilibrium interfacial spacing for each bonding type can be determined. Subsequently, lateral sliding of the slabs allows identification of the most energetically favorable in-plane atomic registry.

As shown in Fig. \ref{interfaceEnergyMap}(a–d), the B–C bonded interface exhibits the lowest interface energy, followed by As–C, with As–Si and B–Si configurations being significantly less stable. The minimum energy interfacial spacings for the B–C and As–C interfaces are approximately 1.5 {\AA} and 1.9 {\AA}, respectively. Although no isolated slab structures with vacuum spacing were included in the training dataset, the DFT single-point calculations of the interface structures agree well with the NEP model predictions (cross symbols in Fig. \ref{interfaceEnergyMap}(a–d)). These results demonstrate the retained predictive capability of the NEP model for surface-related energies even under moderate conditions.

Figures. \ref{interfaceEnergyMap}(e–h) present the energy landscapes associated with lateral sliding at the equilibrium interface distance. Across all four interface types, configurations where the terminating atoms of cBAs and 4H-SiC form staggered, ``egg-box''-like bonding patterns exhibit enhanced energetic stability and greater shear resistance. However, the B–Si and As–Si interfaces show minimal variation in energy over a wide range of interfacial distances (Fig. \ref{interfaceEnergyMap}(b,c)), and their lateral sliding energy profiles lack well-defined local minima (Fig. \ref{interfaceEnergyMap}(f,g)), suggesting that these bonding configurations are unstable or otherwise unfavorable.

Moreover, experimental studies \cite{CHEN200728,C-facejiang2019,C-faceSHEN20237274} have reported that the C-face of 4H-SiC can be readily polished to form an epitaxy-ready surface under high material removal rates, in contrast to the Si-face. This further supports the conclusion that the C-face of 4H-SiC is better suited for integration with cBAs.

\begin{figure*}[htbp]
    \centering \includegraphics[width=1\linewidth,keepaspectratio]{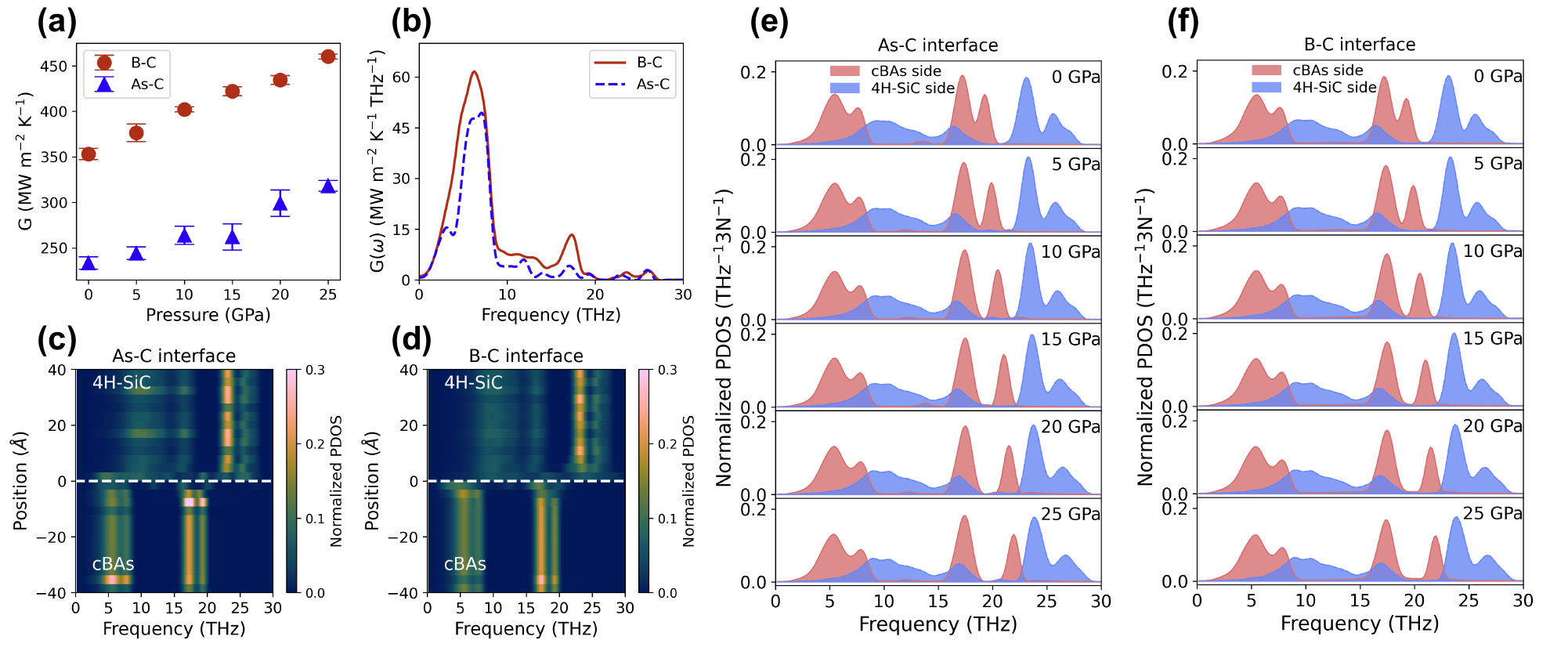}
    \caption{(a) ITC $G$ as a function of applied stress for two bonding configurations (i.e., B–C and As–C bonding) at the cBAs(111)/4H-SiC(0001) interface, obtained from NEMD simulations at 300 K. Error bars represent the standard error from five independent simulations for each data point. (b) Spectral ITC at 300 K, 0 GPa for two type of interface. (c) and (d) Normalized PDOS spatially projected within 40 {\AA} of the As–C and B–C interfaces, respectively. The dashed white line indicates the interface position. (e) and (f) represent the evolution of the PDOS under uniaxial stress for the constituent materials in the As–C and B–C interface models, respectively, calculated in regions on both sides of the interface with a thickness of 40 {\AA}.}
    \label{ITC}
\end{figure*}

\subsection{Uniaxial stress dependence of ITC}

Based on the analysis of interface energies and the relative atomic configurations near the interfaces, a physically reasonable initial structural model was established for performing NEMD simulations. As shown in Fig. \ref{tempprofile}(a), a representative structural model used for NEMD is illustrated. In our simulations, we focused on the two most stable interfaces, namely the B–C and As–C interfaces, as identified from the comparison of interface energy. We first evaluated the ITC of these two interfaces under zero external stress. Taking the B–C interface as an example, Fig. \ref{tempprofile}(b) presents the atomic configuration near the interface, while Fig. \ref{tempprofile}(c) shows the accumulated heat fluxes in the thermostats. The near-perfect symmetry of these fluxes about the zero-energy plane demonstrates compliance with energy conservation, indicating the accuracy of the NEP model and the adequacy of the relaxation time used to establish the steady-state heat flow.

As shown in Fig. \ref{ITC}(a), the As–C interface exhibits a significantly lower ITC than the B–C interface by approximately 120 MW m$^{-2}$ K$^{-1}$ (233 $\pm$ 7 vs. 353 $\pm$ 6 MW m$^{-2}$ K$^{-1}$). This reduction is likely due to the resonance effect arising from the large mass mismatch between interfacial atoms (As:C = 75:12), which inhibits the response of heavy atoms to the ultrafast vibrations of light atoms. The spectrally decomposed thermal conductance $G(\omega)$ in Fig. \ref{ITC}(b) provides further insight. When cBAs and 4H-SiC form an interface, the wide acoustic–optical phonon gap intrinsic to pristine cBAs is partially filled by the broader range of acoustic phonon frequencies in 4H-SiC. Consequently, high-frequency phonons within and beyond the cBAs phonon gap (9–19 THz) contribute more substantially to heat transport. For the B–C interface, phonons above 9 THz account for $\sim$ 20\% of the total ITC, whereas for the As–C interface, this contribution drops to $\sim$ 15\%. This suggests that the large interfacial mass mismatch suppresses the transmission of high-frequency optical phonons, particularly those dominated by the light C atoms in 4H-SiC.

Further evidence comes from the position-projected phonon density of states (PDOS) within 40 {\AA} of the interface region (corresponding to $L_{inter}$=80 {\AA} in Fig. \ref{tempprofile}(a)), shown in Figs. \ref{ITC}(c) and \ref{ITC}(d). In the As–C interface, the PDOS near the interface on the cBAs side displays weaker spectral broadening, and the PDOS on both sides resembles their respective bulk characteristics even near the interface. In contrast, the B–C interface exhibits stronger interaction between the phonons of the two components. To further clarify the phonon characteristics near the two types of interfaces, Figs. \ref{ITC}(e) and \ref{ITC}(f) show the PDOS in a 40 {\AA} region on each side of the interface under different stress conditions. For the As–C interface, a distinct interfacial mode appears in the acoustic–optical phonon gap of cBAs, differing from its bulk PDOS. In comparison, the B–C interface shows a uniform broadening of the PDOS in this frequency range. These interfacial PDOS features diminish at distances beyond 40 {\AA} from the interface (see Fig. S4 \cite{SM}).
\begin{figure}[htbp]
    \centering \includegraphics[width=1\linewidth,keepaspectratio]{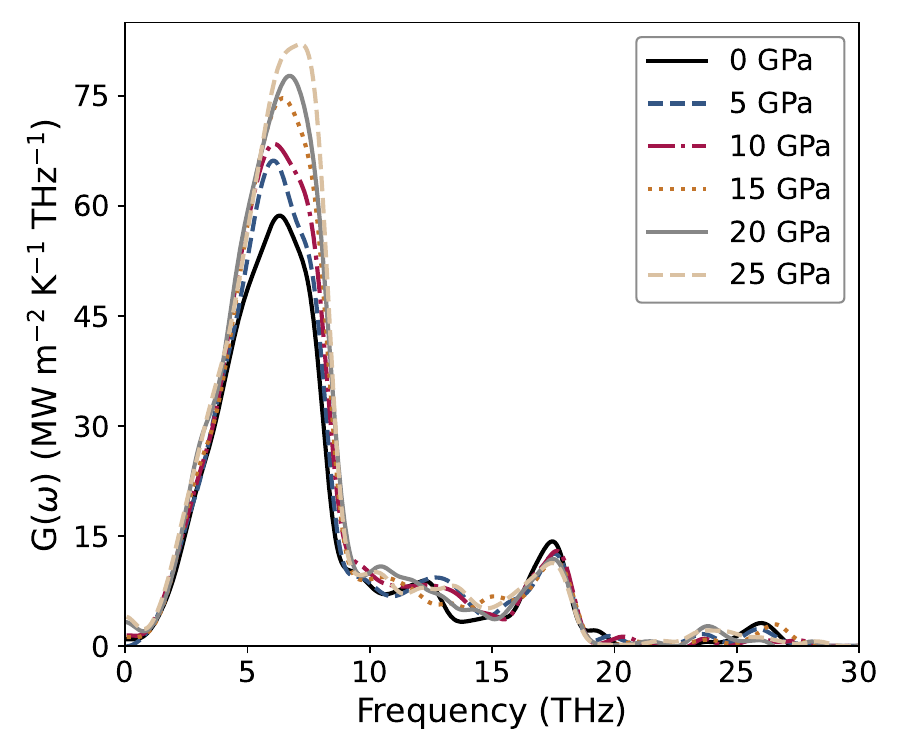}
    \caption{spectrally decomposed ITC $G(\omega)$ of B-C interface under various uniaxial stress.}
    \label{shcOFbc}
\end{figure}
As uniaxial stress increases, the interfacial coupling strengthens, leading to an increase in ITC for both interfaces. This trend can be interpreted by the increase in phonon overlap, quantified by the spectral overlap factor $S$:

\begin{equation}
    S=\frac{\int_{0}^{\infty}D_\text{cBAs}(\omega)D_\text{4H-SiC}(\omega)d\omega}{\sqrt{\left(\int_{0}^{\infty}D_\text{cBAs}(\omega)^{2}d\omega\right) \left(\int_{0}^{\infty}D_\text{4H-SiC}(\omega)^{2}d\omega\right)}},
\end{equation}
where $D_\text{cBAs}$ and $D_\text{4H-SiC}$ represent the phonon density of states (PDOS) of the bulk-like regions of cBAs and 4H-SiC, respectively. The bulk-like regions are defined as slabs with a thickness of 40 {\AA}, positioned 40 {\AA} away from the interface. (see Fig. S4 \cite{SM}). Over the range from 0 to 25 GPa, $S$ increases from 0.20 to 0.26. Specifically, the ITC of the B–C interface increases monotonically with stress, rising from 353 $\pm$ 6 MW m$^{-2}$ K$^{-1}$ at 0 GPa to 460 $\pm$ 3 MW m$^{-2}$ K$^{-1}$ at 25 GPa. Similarly, the As–C interface shows an increase from 233 $\pm$ 7 to 318 $\pm$ 6 MW m$^{-2}$ K$^{-1}$ over the same stress range.

To elucidate the stress-dependent phonon contributions to the ITC of the B–C interface, the Fig. \ref{shcOFbc} presents the spectral ITC $G(\omega)$ under various uniaxial stresses. With increasing uniaxial stress, the relative contribution of the acoustic phonon frequency range to the ITC further increases. Meanwhile, the corresponding frequency range becomes broader overall, and the frequency at which the maximum contribution occurs shifts toward higher values. This behavior aligns with the stress-induced hardening of acoustic phonon modes (see Fig. S5 \cite{SM}). In contrast, the contribution from higher-frequency optical phonons (i.e. $ \sim$ 9 – 19 THz ) remains nearly unchanged with increasing stress.

\section{Conclusion}

To systematically explore the ITC and its stress response at the cBAs(111)/4H-SiC(0001) heterointerface, we developed a NEP model that accurately captures the atomic interactions in 4H-SiC, high-$\kappa$ cBAs, and their interface. Following thorough validation, we confirmed that the B–C bonded configuration corresponds to the lowest-energy interface for cBAs(111)/4H-SiC(0001), while the As–C bonded configuration represents a metastable alternative. The optimal interfacial atomic stacking arrangement was also determined. Using the most stable interfacial configuration as input for NEMD simulations, we systematically examined the variation of ITC with increasing uniaxial stress at 300 K for both B–C and As–C bonded interfaces. Our results show that the reduced ITC of the As–C interface originates from the significant atomic mass mismatch at the interface, which hinders phonon transmission. Moreover, the ITC of both interface types increase monotonically with uniaxial stress, from 353 $\pm$ 6 to 460 $\pm$ 3 MW m$^{-2}$ K$^{-1}$ for B–C and from 233 $\pm$ 7 to 318 $\pm$ 6 MW m$^{-2}$ K$^{-1}$ for As–C, reflecting strengthened interfacial bonding under compression. These findings underscore the potential of cBAs as a promising substrate for thermal management in power electronics. In particular, when integrated with 4H-SiC, the ITC across the cBAs/4H-SiC interface can be significantly modulated by perpendicular uniaxial stress. Our work provides an atomistic perspective for understanding and optimizing the thermal performance of such heterojunctions, suggesting that moderate stress engineering may yield substantial enhancements in ITC.

\section*{DATA AVAILABILITY}

The NEP model used in this study for investigating the ITC of the cBAs(111)/4H-SiC(0001) interface, along with the associated DFT reference data, is openly available in the GitHub repository at \url{https://github.com/Lenslike/publication/tree/main/cBAs-4H-SiC_Dataset}.

\section*{Acknowledgements}

This work was supported by the National Key R{\&}D Program of China (Grant No. 2023YFB4603801), National Natural Science Foundation of China (Grant Nos.12474249), Guangdong Provincial Key Laboratory of Magnetoelectric Physics and Devices (No. 2022B1212010008).

\nocite{comparisonSED,ThomasSED,EqualityofBothSED,larkin2014comparisonSED,liang2025pysedtool,Anomalous-pressure-dependenceofBAs,Nonmonotonic-P_dependOFBAs,li2022anomalous,elasticConstantsof4HSiC,elasticConstantsofc-BAs,cBAshardness}

\bibliography{cite}
\end{document}